\documentclass[twocolumn,showpacs,preprintnumbers,amsmath,amssymb]{revtex4}
\usepackage{graphicx}
\usepackage{dcolumn}
\usepackage{xcolor}	
\newcommand{\degree}{\ensuremath{^\circ}}			

\newcommand{\of}[1]{\ensuremath{(#1)}}					
\newcommand{\OF}[1]{\ensuremath{\left(#1\right)}}			
\newcommand{\OFbrace}[1]{\ensuremath{\left\lbrace#1\right\rbrace}}	
\newcommand{\pd}[2]{\ensuremath{\frac{\partial #1}{\partial #2}}}	

\newcommand{\rt}[1]{\ensuremath{\tilde{\rho}_{#1}}}		
\newcommand{\vt}[1]{\ensuremath{V_{#1}}}			
\newcommand{\rtk}{\ensuremath{\tilde{\rho}_{k}}}		
\newcommand{\vtk}{\ensuremath{V_{k}}}				
\newcommand{\Qr}{\ensuremath{Q}}				
\newcommand{\qr}{\ensuremath{q}}				
\newcommand{\Qi}{\ensuremath{B}}				
\newcommand{\qi}{\ensuremath{\beta}}				
\newcommand{\Rshell}{\ensuremath{R_\text{S}}}			

\newcommand{\rone}{\ensuremath{r_{\text{\tiny{I}}}}}		
\newcommand{\rtwo}{\ensuremath{r_{\text{\tiny{II}}}}}		
\newcommand{\rthree}{\ensuremath{r_{\text{\tiny{III}}}}}	
\newcommand{\gangle}{\ensuremath{\vartheta}}			
\newcommand{\gangleS}{\ensuremath{\varphi}}			
\newcommand{\gangleone}{\ensuremath{\gangle_{\text{\tiny{I}}}}}	
\newcommand{\gangletwo}{\ensuremath{\gangle_{\text{\tiny{II}}}}}	

\newcommand{\done}{\ensuremath{d_{\text{\tiny{I}}}}}		
\newcommand{\dtwo}{\ensuremath{d_{\text{\tiny{II}}}}}		

\newcommand{\rdelta}{\ensuremath{\Delta l}}			
\newcommand{\pdelta}{\ensuremath{\Delta p}}			

\newcommand{\kB}{\ensuremath{k_\text{B}}}			
\newcommand{\cheat}{\ensuremath{c_\omega}}			

\renewcommand{\vec}[1]{\ensuremath{\mathbf{#1}}}		

\begin{document}
\bibliographystyle{apsrev}


\title{Resolving structural transitions in spherical dust clusters}

\author{H.~Thomsen$^1$}
\author{M.~Bonitz$^1$}
\affiliation{$^1$Institut f\"ur Theoretische Physik und Astrophysik, Christian-Albrechts-Universit\"{a}t zu Kiel, D-24098 Kiel, Germany}


\date{\today}
\begin{abstract}
Finite systems in confining potentials are known to undergo structural transitions similar to phase transitions. However, these systems are inhomogeneous, and their ``melting'' point may depend on the position in the trap and vary with the particle number. Focusing on three-dimensional Coulomb systems in a harmonic trap a rich physics is revealed: in addition to radial melting we demonstrate the existence of intrashell disordering and inter-shell angular melting. Our analysis takes advantage of a novel melting criterion that is based on the spatial two and three-particle distribution functions and the associated reduced entropy which can be directly measured in complex plasma experiments.
\end{abstract}

\maketitle
\section{Introduction}
Sudden changes of the properties of a many-particle system upon continuous change of its parameters are among the most fascinating phenomena in nature. Such phase (or structural) transitions have been discovered not only in physics but also in chemistry, biology and even social systems. Experimental detection or theoretical prediction of the transition point and analysis of its properties (such as the critical exponents, e.g. \cite{critical_indices}) is of fundamental importance for understanding the underlying physics and common features in, possibly, entirely different systems.
In macroscopic systems suitable quantities to pinpoint the phase transition 
are based on the free energy, on energy fluctuations (heat capacity, $c_V$) or on particle position fluctuations (e.g. Lindemann criterion). 
Microscopic approaches link the melting point to the interparticle correlations in the system, e.g. to a characteristic peak height of the static structure factor \cite{hansen69} or a jump of the first peak height of the pair distribution function \cite{ott_pop}.
As a Lindemann-like melting criterion, the fluctuations of the bondlength can quantify the level of rigidity in a finite cluster~\cite{calvo_melting_2007}.
Other concepts are based on transport properties such as diffusion \cite{loewen_melting}. 

While phase transitions pertain to macroscopic systems only, solid-like or liquid-like behavior has been observed in finite systems containing as few as $10$ particles, e.g. in quantum dots~\cite{bedanov, filinov-etal.01prl}, ions in traps \cite{itano}, dusty plasma crystals \cite{bonitz-etal.prl06}, atomic clusters \cite{frantz,proykova06} and polymers \cite{berry02} etc. The notion of liquid and solid ``phases'' has been used successfully to characterize qualitatively different behaviors which resemble the corresponding properties in macroscopic systems, for a further discussion, see \cite{boening-prl08}.

The melting process is much more complex in finite systems that are attracting growing interest in recent years---interfaces \cite{dosch_04}, two-dimensional layers \cite{sheridan08, goree10,nosenko13}, gas or metal clusters, e.g. \cite{baletto05,stamerjohanns_clusters_2002}, trapped ions \cite{itano}, ultracold atoms or molecules in traps, dust \cite{bonitz_rpp10, hartmann_prl10, calvo_melting_2007} or colloidal clusters \cite{ivlev_loewen} or electrons in quantum dots \cite{filinov-etal.01prl, ghosal06}.
Metal clusters have been subject to several experimental studies, e.g.~\cite{buffat_size_1976,lai_melting_1998,dippel_size-dependent_2001}, theoretical works, e.g.~\cite{reiss_capillarity_1988,vanfleet_thermodynamics_1995} and to numerical simulations, e.g.~\cite{qi_melting_2001,ding_size_2004}.
A concordant result is that the stability of these clusters with short-range interaction increases with the cluster size. Smaller clusters exhibit a melting point depression proportional to one over the cluster radius.
A similar behavior is observed in finite clusters with long-range Coulomb interaction (ions in traps, colloids).
Schiffer found in simulations a decrease of the melting temperature compared to a bulk system, $T_\text{m}\of{N}-T_\text{m}^{\infty} \propto -N^{-1/3}$, which was attributed to the fraction of particles in the surface layer that scales as $\propto N^{-1/3}$~\cite{schiffer_melting_2002}.

Finite clusters of dust particles trapped in a plasma discharge are special representatives of such mesoscopic systems for which many details of structural transitions can be studied experimentally in unprecedented detail. This is due to the possibility to track the position and velocity of each individual particle. Therefore, in the following we will concentrate on finite dust clusters although many of the results are expected to be applicable to other finite systems as well. In particular, when discussing quantities that are suitable for the analysis of phase transitions we will concentrate on those that are accessible in experiments.

A characteristic feature of spherical dust clusters is that their density is inherently inhomogeneous, e.g. \cite{bedanov, filinov-etal.01prl},
and their melting point may be space-dependent or depend on the precise particle number $N$, e.g.~\cite{schiffer_melting_2002, haberland_melting_2005, vova_jpa06} and on ``magic'' configurations (closed shells).
Furthermore, the melting process of small Coulomb clusters is known to consist of multiple transition~\cite{apolinario,ogawa_two-step_2006,apolinario_2007,apolinario14}.
This makes a theoretical analysis of melting complicated because different quantities that agree among each other for macroscopic systems may yield very different predictions in finite systems. For example,
Stamerjohanns \textit{et al.} showed that certain transitions in finite clusters are not captured by the specific heat capacity~\cite{stamerjohanns_clusters_2002}.

A prominent example of finite clusters are repulsively interacting particles in a harmonic trap that, at low temperature, are localized on concentric rings (in two dimensions, 2D) or shells (in 3D), cf. Fig.~\ref{fig:3d-sketch}. These are crystal-like clusters with complex spatial correlations of particles within and between shells. In 2D systems, temperature increase leads to a peculiar two-step ``melting'' \cite{bedanov,filinov-etal.01prl, schablinski_vidf}: first the inter-shell angular correlations are lost (shells are free to rotate against each other) and, at a higher temperature, particle transitions between shells set in eventually destroying the shell structure (``radial melting'').
Especially the former process depends crucially on the exact occupation numbers of the shells.

In 3D clusters the situation is more complex. Here inter-shell transitions have been observed in experiments \cite{block8, schella11} and explored theoretically \cite{vova_jpa06, apolinario,hanno_pre8}.
There have been some predictions about additional melting-type processes, e.g. \cite{apolinario,wrighton12}. In contrast, a recent analysis based on the diffusivity of finite systems \cite{melzer-inm} showed just one structural transition, in agreement with experiments on laser heated dust clusters \cite{thomsen_jpd14}. This is surprising since there is no obvious reason why the intrashell disordering within shells should occur simultaneously with radial melting or why 3D clusters should behave qualitatively different from 2D ones. We will show below that the main reason is that, many quantities (such as diffusion coefficients) studied so far, are not suitable to distinguish between different types of structural transitions 

It is the purpose of this paper to resolve this problem. Here we show that the physics of 3D clusters with long range interaction is much richer than observed in experiments so far, in agreement with theoretical studies, e.g. \cite{apolinario, ogawa_two-step_2006}. 
In fact, several structural transitions exist:
\begin{description}
 \item[i.] Radial melting (RM)---this describes the process in which the radial order between distinct shells is lost. 
 \item[ii.] Intrashell disordering (ID)---the loss of angular order within one shell (this is closely related to 
angular melting \cite{bedanov, filinov-etal.01prl, apolinario}) and may occur in different shells at different temperature. 
Typically, we observe that this is a gradual process that is very sensitive to the precise particle number. At low temperatures often transitions between different {\em intrashell isomers} (without changing the particle number within the shell) are observed, yet the angular order is lost completely only during the radial melting process.
 \item[iii.] Inter-shell melting (ISM)---an angular disordering process that is well known from 2D systems where, below a critical temperature, the relative angle of two adjacent shells is fixed and the shells are ``locked''.
A similar angular locking is possible for 3D shells as well, but we observe it only at very low temperature. 
\end{description}
One driving mechanism that accompanies all of the above structural transitions are inter-shell particle transitions (sometimes referred to as transition between different radial isomers, i.e. configurations with different shell occupations). Inter-shell transitions usually set in in at a temperature well below that of RM where distinct shells still exist. As we will see, inter-shell transitions also affect the intrashell disordering.
We will show, based on first principle Monte Carlo (MC) simulations, that these structural transitions may be substantially displaced from each other in temperature. We predict the characteristic values of the critical temperatures, for various typical clusters so that an experimental observation should be possible.

Our analysis is based on a novel general melting criterion that is derived from the reduced entropy and the associated reduced heat capacity of two and three particle complexes. A particularly useful feature is that it is computed exclusively from the (correlated) particle positions which are directly accessible in dusty plasma experiments.

This paper is organized as follows: In Sec.~\ref{s:theory} we introduce the reduced $k$-particle densities for spatially inhomogeneous spherically trapped clusters and define the associated entropies $S^{(k)}$ and heat capacities $c^{(k)}$. 
A first test is then performed for an infinite homogeneous 2D layer where the peak of $c^{(2)}$ precisely agrees with the known melting point. After specializing $S^{(k)}$ and $c^{(k)}$ to spherically confined systems, we present, in Sec.~\ref{s:results}, numerical results for three representative spherical dust clusters that are summarized in Sec.~\ref{s:conclusion}.

\section{Theory}\label{s:theory}
\subsection{Theoretical basis: Spatial two- and three-particle correlation functions}
The correlations of the particle positions and the structural order in a macroscopic system are characterized by the radial pair distribution function (PDF) $g\of{r_{12}}=g\of{|\vec{r}_2 - \vec{r}_1|}$---the probability to observe an arbitrary pair of particles at a distance $r_{12}=|\vec{r}_{12}|$. The dependence on the modulus ${r}_{12}$ alone is a consequence of the translational and rotational invariance of the Hamiltonian. In a crystalline state the latter is broken and, if the system is finite, also the former symmetry is lost. Then the natural generalization is the full two-particle distribution function $\rho_2\of{\vec{r}_1,\vec{r}_2}$---the joint probability to observe one particle at an exact space point $\vec{r}_1$ and a second one at $\vec{r}_2$. Even this may be not sufficient to uniquely describe the local structure, in particular, in case of particle ordering on spherical shells, as it occurs in traps. In that case, a more sensitive quantity is the three-particle distribution function $\rho_3\of{\vec{r}_1,\vec{r}_2,\vec{r}_3}$. We will show below that $\rho_2$ and $\rho_3$ are very well suited to quantitatively study the order and structural changes upon variation of temperature, and we also present a simple approach how to detect from them the point(s) of structural transitions. 

The classical $k$-particle equilibrium distribution ($k=1\dots N$) follows from the full $N$-particle distribution $\rho$ via
integration over the remaining $N-k$ positions~\cite{nettleton_entropy_1958,hansen_textbook},
\begin{align}
 \rho_k\of{\vec{r}_1,\dots,\vec{r}_k} &=
  \frac{1}{(N-k)!}
  \nonumber\\
  &\cdot \int{\text{d}^3 r_{k+1}} \hdots \int{\text{d}^3 r_N} \
  \rho\of{\vec{r}_1,\hdots,\vec{r}_N}
  \label{eq:k_part_density} \text{ ,}
\end{align}
where $\rho$ is normalized to $N!$.
We introduce a new set of coordinates, $(\Qr,\Qi)$ with dim $\Qr=\qr$, dim $\Qi=\qi$ and $\qr+\qi=3k$, where $\rho_k$ is independent 
of all components of $\Qi$, as a consequence of symmetries of the Hamiltonian. So in the new variables 
$\rho_k \to \rtk$ depending only on $\qr$ arguments.
The associated coordinate transformation is defined by $\left(\vec{r}_1,\dots,\vec{r}_k\right)=\Phi\OF{\Qr,\Qi}$, and 
$\left|\mathcal{J}\Phi\OF{\Qr,\Qi}\right|$ is the Jacobi determinant.
%
The distribution function in the new variables is
\begin{align}
 \rtk\OF{\Qr} &= \underbrace{\int{\text{d}^{\qi} \Qi} \left|\mathcal{J}\Phi\OF{\Qr,\Qi}\right| }_{:=\vtk\of{\Qr}}
  \underbrace{\rho_{k}[\Phi\OF{\Qr,\Qi}]}_{\text{independent of } \Qi} 
  \label{eq:k_part_density_generalized_02}. \\
&= \vtk\of{\Qr} \cdot \rho_{k}[\Phi\OF{\Qr,\Qi}]
  \label{eq:k_part_density_generalized_04}\text{ .}
\end{align}
Note that $\rtk\OF{\Qr}$ will be practically sampled from the coordinates obtained in experiments or numerical simulations.
However, the functional form of the geometrical factor $\vtk$ is important for the normalization and for derivations of the reduced entropy $S^{(k)}$.
Following the general introduction of the $k$-particle density, this concept is first explained for an infinite 2D system before proceeding to the system of our main interest---spherical 3D clusters.

\subsection{Test case: Macroscopic 2D Yukawa system}\label{ss:2d}
Before applying $\rt{2}$ and $\rt{3}$ in generalized coordinates to the spherical system~(\ref{eq:h}), we test them for a well studied case of an {\bf infinite 2D layer (Yukawa OCP)}.
Therefore, the general result~(\ref{eq:k_part_density_generalized_04}) is first applied to this system for the pair distribution function (PDF, $k=2$) and the triple correlation function (TCF, $k=3$).

\begin{itemize}
 \item For the PDF, we use Cartesian coordinates for the first position and  polar coordinates ($r_{12}, \varphi$) for the distance between particle one and two. 
 Due to the translational and rotational invariance of the Hamiltonian, we have $\Qr=\lbrace r_{12} \rbrace$ 
 and $\Qi=\lbrace x_1, x_2, \varphi \rbrace$.

 The Jacobian is $\left|\mathcal{J}\Phi\OF{\Qr,\Qi}\right| = r_{12}$, whereas 
 the geometrical factor in the pair distribution becomes
 \begin{align}
  \vt{2}\of{r_{12}} &= 
  \int{\text{d}^2 r_1} \int_{0}^{2\pi}{\text{d}\varphi}\ r_{12}
  = 2\pi r_{12} V
  \label{eq:pdf_vtilde_01} \text{ ,}
 \end{align}
 where $V$ is the volume of the system.
 In order to be consistent with the textbook definition of the PDF, this geometrical factor is multiplied by the squared overall density $\rho_{0}^{2}$ in
 \begin{align}
  \rt{2}\of{r_{12}} &= \text{PDF}\of{r_{12}} \cdot 2\pi r_{12} V \cdot \rho_{0}^{2} \text{ .}
  \label{eq:PairDistFunction_def}
 \end{align}
 \item For the TCF in the flat 2D system, the first position is again described in Cartesian coordinates.
 We introduce polar coordinates with the $x$-axis along $\vec{r}_{2}-\vec{r}_{1}$ for describing the second and the third position relatively to the first position. 
 The latter two position vectors are $\vec{r}_2 \to \lbrace \done, \varphi_1 \rbrace$ 
 and $\vec{r}_3 \to \lbrace \dtwo, \gangleS \rbrace$, and we have  
  $\Qr=\lbrace \done, \dtwo, \gangleS \rbrace$ and 
  $\Qi=\lbrace x_1, x_2, \varphi_{1} \rbrace$. 
  The Jacobian is $\left|\mathcal{J}\Phi\OF{\Qr,\Qi}\right| = \done \dtwo$, and the geometrical factor in the TCF (2D) becomes
 \begin{align}
  \vt{3}\of{\done,\dtwo,\gangleS} &= 
  2 \cdot \int{\text{d}^{2} r_1} \int_{0}^{2\pi}{\text{d}\varphi_1} \ \done \dtwo
 \label{eq:tcf_flat_vtilde_00}\\
   &= 4\pi \done \dtwo V \hspace{6em} \text{(2D)}
  \label{eq:tcf_flat_vtilde_01} \text{ ,}
 \end{align}
where the factor $2$ in front the integral has to be added, since we do not resolve the orientation of the bond angles.
Bond angles above $\gangleS'>\pi$ are mapped to the interval $[0:\pi]$ by $\gangleS' \mapsto 2\pi-\gangleS'$. 
The TCF (2D) is defined by
\begin{align}
 \rt{3}\of{\done,\dtwo,\gangleS} &= \text{TCF}\of{\done,\dtwo,\gangleS} \cdot 4\pi \done \dtwo V
 \label{eq:tcf_flat_definition} \text{ ,}
\end{align}
as an extension of the PDF to particle triples.
\end{itemize}
To analyze structural (``melting-like'') transitions upon heating it is advantageous to introduce a dimensionless {\em coupling parameter}---the ratio of the mean interaction energy of neighboring particles to their thermal energy--$\Gamma = \frac{e^2}{a \kB T}$. For Yukawa interaction the coupling parameter has to be modified as discussed e.g. in Ref. \cite{ott_pop} but this is not essential for the present discussion.
As a result the system state dependence (the ensemble) is reduced to three parameters $(N, \kappa, \Gamma)$,

A typical example of the TCF (2D) for a solid and liquid system is shown in Fig.~\ref{fig:2d}(a,b), respectively, indicating a 
strong temperature dependence of the TCF.

\subsection{Reduced entropies.}
What remains is to find a quantitative measure that allows one to detect, from the TCF (or PDF), the ``phase'' boundary and the ``melting'' point(s).
Here we propose to use the Shannon entropy (or negative information) \cite{shannon1948} which can be interpreted as a measure for the disorder (or information) of a statistical distribution. Before presenting our definitions and results we briefly recall similar approaches that are based on reduced distribution functions.

The entropy concept has been of fundamental importance in nonequilibrium in the context of irreversibility of relaxation processes (Boltzmann's H-theorem \cite{boltzmann}) and also for open and chaotic systems, e.g. \cite{klimontovich1, klimontovich2}.
On the other hand, the idea to calculate, for a correlated equilibrium system, a reduced entropy from $k$-particle distributions has been widely used for extended systems~\cite{nettleton_entropy_1958, raveche_entropy_1971, mountain_entropy_1971}, especially in the context of glass transitions~\cite{mittal_liquids_2006,kawasaki_glass_2007}. The first new point in our work is that we present a result that is applicable to spatially inhomogeneous systems.

An early idea is due to Stratonovich \cite{stratonovich_entropy}
who defined separate entropy-type contributions from correlated configurations of $2, 3, \dots$ particles. While this certainly captures all correlation effects in the system, individual contributions carry a high degree of redundancy since, in most cases two- and three-particle correlations are sufficient to characterize the many-particle state. We will, therefore, concentrate on the reduced entropies computed from either the two-particle or the three-particle distribution function alone. Thus, we are not attempting to reproduce the exact thermodynamic entropy but to capture the dominant correlation contributions and their temperature dependence.
Finally, while earlier works often studied the entropy for certain analytical approximations for the higher order distribution functions, e.g. \cite{nettleton_entropy_1958, raveche_entropy_1971} and references therein, we will avoid any approximation, but use the exact pair and three-particle distribution calculated from a first-principle computer simulation.

We now proceed with our approach to the reduced entropy as a measure for the structural order of correlated equilibrium systems. We will first study the example of a macroscopic 2D system of Sec.~\ref{ss:2d} and generalize the result to 3D clusters in Sec.~\ref{ss:entropy-3d}.
Since $\rho_2$ ($\rho_3$) is a well-defined, normalized and non-negative probability density it can be used to compute averages of all observables that depend on not more than two (three) coordinates. A suitable observable is then $-{\rm ln}\rho_2$ ($-{\rm ln}\rho_3$) which yields the reduced two-(three-) particle entropy [in units of $k_B$]. 
The spatial contribution of the well-known thermodynamic entropy of $N$ classical particles in 2D can be expressed as
\begin{align}
 S &= -\frac{\kB}{N!} \int{\text{d}^2 \vec{r}_1} \hdots \int{\text{d}^{2} \vec{r}_N} \ 
  \rho\of{\vec{r}_1,\hdots,\vec{r}_N}
  \nonumber\\
  &\hspace{4em}\cdot \ln\OFbrace{\rdelta^{2\cdot N}\rho\of{\vec{r}_1,\hdots,\vec{r}_N}}
  \label{eq:spatial_entropy} \text{ ,}
\end{align}
where $\rho$ is the spatial $N$-particle distribution with normalization $N!$ and 
$\rdelta$ results from the decomposition of the phase-space cell into spatial and
momentum width $2\pi \hbar = \pdelta \cdot \rdelta$.
The actual choice of $\rdelta$ just causes a constant offset in $S$.

In analogy with Eq.~\eqref{eq:spatial_entropy}, we introduce the reduced entropy of the three-particle density as
\begin{align}
 S^{(3)} &= -\kB \frac{(N-3)!}{N!} \int{\text{d}^2 \vec{r}_1} \int{\text{d}^2 \vec{r}_2} \int{\text{d}^2 \vec{r}_3} \
 \rho_{3}\of{\vec{r}_1,\vec{r}_2,\vec{r}_3} 
 \nonumber\\
 &\hspace{8em}\cdot \ln\OFbrace{\rdelta^{6}\rho_{3}\of{\vec{r}_1,\vec{r}_2,\vec{r}_3}}
 \label{eq:three_part_entropy} \text{ .}
\end{align}
Without any further approximation, we can execute this integration in generalized coordinates and apply Eq.~\eqref{eq:k_part_density_generalized_04},
\begin{align}
 S^{(3)} &= -\kB \frac{(N-3)!}{N!} \int{\text{d}^\qr \Qr} 
  \underbrace{ \int{\text{d}^{\qi} \Qi} \left|\mathcal{J}\Phi\OF{\Qr,\Qi}\right| }_{=\vt{3}\of{\Qr}}
  \nonumber\\
  &\hspace{4em}\cdot\underbrace{
  \frac{\rt{3}\of{\Qr}}{\vt{3}\of{\Qr}}
  \ln\OFbrace{\rdelta^6 \frac{\rt{3}\of{\Qr}}{\vt{3}\of{\Qr}}}
  }_{\text{independent of }\Qi}
  \label{eq:three_part_entropy_generalized_01}\\ &
  = -\kB \frac{(N-3)!}{N!} \int{\text{d}^\qr \Qr}\ 
    \rt{3}\of{\Qr} \ln\OFbrace{\rdelta^6 \frac{\rt{3}\of{\Qr}}{\vt{3}\of{\Qr}}}
  \label{eq:three_part_entropy_generalized_02} \text{ ,}
\end{align}
with $\Qr=\left\lbrace \done, \dtwo, \gangleS \right\rbrace$ and $\vt{3}=4\pi \done \dtwo$ for the TCF (2D).
$\vt{3}$ in the denominator cancels the Jacobi determinant under the integral, but under the logarithm  $\vt{3}$ remains.
Formally, we can write the reduced entropy as the expectation value of the logarithm of the PDF or the TCF, respectively, in the canonical ensemble,
\begin{align}
    S^{(2)} &\equiv -\left\langle \ln\text{PDF} \right\rangle \text{, }
    &
    S^{(3)} &\equiv -\left\langle \ln\text{TCF} \right\rangle
 \label{eq:entro_2D}	\text{ ,}
\end{align}
where $\langle \dots \rangle$ denotes averaging with $\rt{2}$ and $\rt{3}$, respectively.
The $\rdelta$ factor can be absorbed by the system of dimensionless units.
The derivation for the reduced entropy $S^{(2)}$ associated with the two-particle density $\rho_2$ is analogous.
\begin{figure}[t]
 \includegraphics{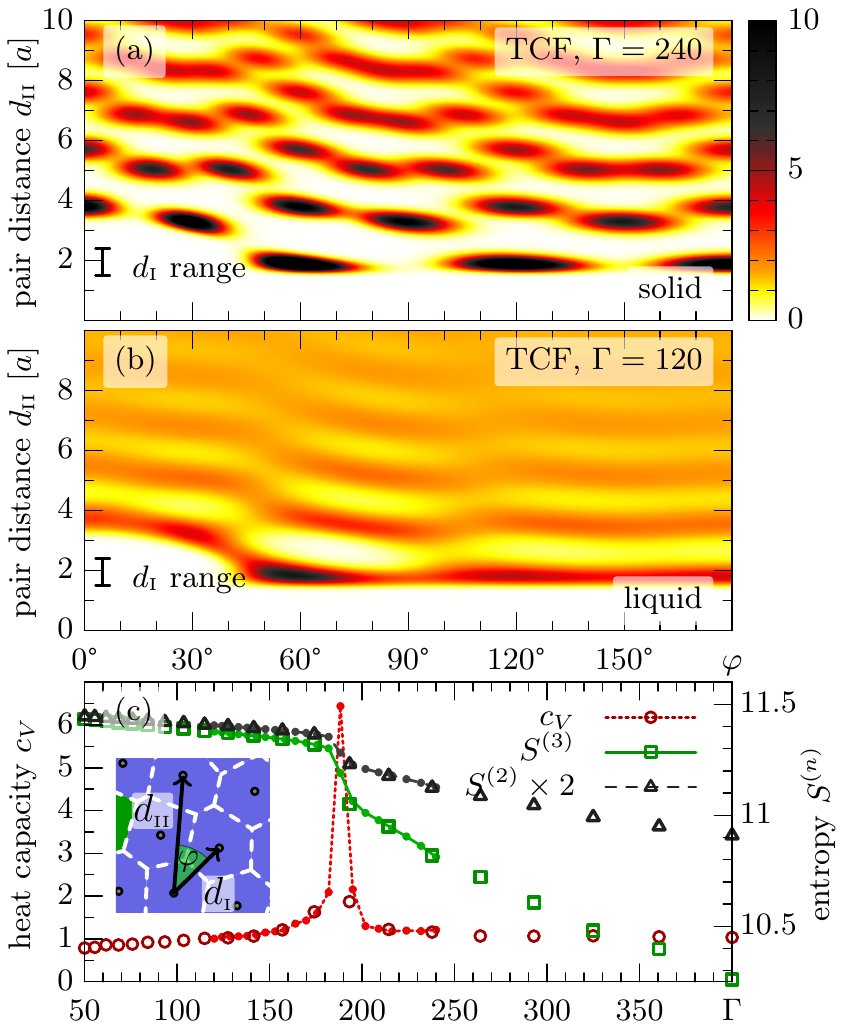}
 \caption{Macrosopic 2D Yukawa OCP, $\kappa=1$, $N=2000$: 
    \textbf{(a)} TCF in the crystal phase.
    \textbf{(b)} TCF in the strongly correlated liquid regime.
   \textbf{(c)} Specific heat $c_V$ and reduced entropies of TCF and PDF, Eq.~(\ref{eq:entro_2D}).
    Inset: sketch of TCF parameters for a 2D system. In (a) and (b) the TCF is averaged over a finite $\done$ 
    range (cf. bars). In (c) no averaging is applied to $S^{\rm (3)}$. 
  }
\label{fig:2d}
\end{figure}

We perform parallel tempering MC simulations, see below, with $N=2000$ particles and periodic boundary conditions and sample the functions (\ref{eq:PairDistFunction_def}, \ref{eq:tcf_flat_definition}) from the particle positions.
We sample 3D histograms from which we compute $\rt{2}$ and $\rt{3}$. 
We verified convergence with respect to the number $M$ of discretization cells, typically this is achieved for $M=300 \dots 400$ for each direction. 
Then, we compute $S^{(2,3)}$ according to Eq.~(\ref{eq:entro_2D}).

The TCF for $\Gamma$ in the solid and liquid regimes is shown in Fig.~\ref{fig:2d}(a) and (b), respectively.
Since we can plot the TCF only as a function of two coordinates, we average over the argument $\done$, selecting a finite range around the nearest neighbor distance (see figures).
The hexagonal order in the solid regime manifests itself in preferred bond angles $\gangleS$ at multiples of ${60}{\degree}$ [${30}{\degree}$] for nearest neighbors at $\dtwo \approx 2$  [for second neighbors at $\dtwo\approx4$].

Finally, the entropies $S^{(2,3)}$ are plotted versus $\Gamma$ in Fig.~\ref{fig:2d}(c). 
They exhibit a sharp drop at $\Gamma\approx185$---at the peak of the specific heat $c_V$ which is obtained in the same simulation.
This is just the freezing point known from the literature \cite{hartmann_2d,ott_pop}.

This behavior suggests to analyze the quantity
\begin{align}
 c_V^{(k)} \equiv -\left.\frac{\partial S^{(k)}}{\partial \ln \Gamma}\right|_V = T \left.\frac{\partial S^{(k)}}{\partial T}\right|_V, \quad k = 1, 2, 3, \dots,
\label{eq:cvi_def}
\end{align}
where $(k)$ refers to the $k$-particle distribution function used in Eq.~(\ref{eq:entro_2D}).
In our case, $c_V^{(2)}$ and $c_V^{(3)}$ refer to the PDF and TCF, respectively.
The physical reason for the good agreement between the peaks of $c_V$ and $c_V^{(2,3)}$ is obvious: equation~(\ref{eq:cvi_def}) coincides with the definition of the specific heat (heat capacity in units of $k_B$ and $N$) of a thermodynamic (infinite) system, provided the thermodynamic entropy $S$ (computed from the full canonical Gibbs distribution) is substituted for $S^{(k)}$. If the local order in the system is dominated by pair (three-particle) correlations then the reduced entropy $S^{(2)}$ ($S^{(3)}$) should capture the temperature dependence of $S$, and the ``reduced specific heat'' $c_V^{(2)}$ ($c_V^{(3)}$) should reproduce $c_V$.

\subsection{Reduced entropies of charged particles in a 3D harmonic trap}\label{ss:entropy-3d}
\begin{figure}
 \includegraphics[width=50mm]{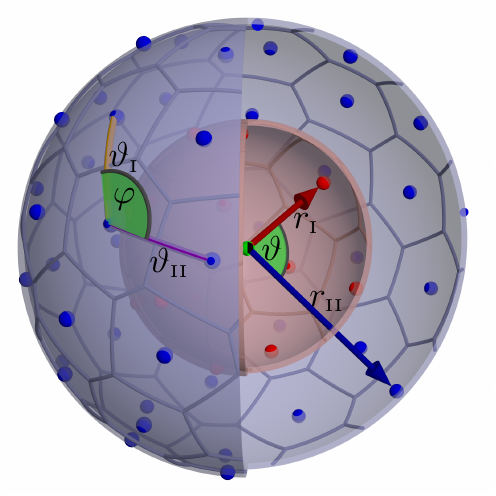}
 \caption{(Color online) Sketch of a spherical 3D cluster with two shells [particles on inner and outer shell are drawn in red and blue, respectively]. Intrashell Voronoi diagrams (nearest neighbors of enclosed particles) are sketched in gray. In the C2P, the radii $\rone$ and $\rtwo$ as well as the pair angle $\gangle$ are recorded (right). In the TCF three particles are chosen from the same shell (left) and two angular pair distances $\gangleone$, $\gangletwo$ and the bond angle $\gangleS$ are recorded. 
}
 \label{fig:3d-sketch}
\end{figure}
We now return to the system of interest---charged particles trapped in an isotropic 3D harmonic trap described by the Hamiltonian
\begin{equation}
	\hat H = \sum_{i=1}^N \frac{\vec p_i^2}{2m} + \sum_{i=1}^N \frac{m}{2}\omega^2 \vec r_i^2 +
                 e^2\sum_{1\le i<j}^N \frac{e^{-\kappa r_{ij}}}{r_{ij}}, 
	\label{eq:h}
\end{equation}
where $r_{ij} = |{\vec r}_i - {\vec r}_j|$. The particles have a pairwise repulsion which includes long-range (Coulomb, $\kappa=0$) interaction and short-range Yukawa-Debye interaction with the inverse screening length $\kappa$.
The model (\ref{eq:h}) has been very successful in describing trapped particles in many fields, including electrons in quantum dots and ions in traps ($\kappa=0$) as well as colloidal systems and complex plasmas ($\kappa a\sim (0.5\dots 2)$, where $a$ is the Wigner-Seitz radius, e.g. \cite{bonitz-etal.prl06, apolinario}). 
The ground state of this system consists of concentric spherical shells (3D), cf. Fig.~\ref{fig:3d-sketch}, and is well understood from simulations~\cite{itano,schiffer_melting_2002, ludwig05}.
In the following the system (\ref{eq:h}) is considered to be in a heat bath with temperature $T$ and fixed particle number $N$ and trap parameters, i.e. in a modified canonical ensemble ($N, \omega, T$) where the role of the volume is taken over by the trap frequency $\omega$ that controls the mean density. All quantities are now computed at fixed $\omega$ instead of volume $V$. 

In the definition of the coupling parameter $\Gamma$, the length $r_{0}=\sqrt[3]{e^2/m\omega^2}$ is used as a representative of the Wigner-Seitz radius for spherical clusters.
As shown in Ref.~\cite{henning_2007}, this length has the correct limit $r_{0}=a$ for large Coulomb balls. 
In the following $r_{0}$, $E_{0}=(e^2/(m\omega^2))^{1/3}$ and $t_{0}=\omega^{-1}$ are used as units for length, energy and time, respectively.
By the application of the coupling parameter, the system state dependence (the ensemble) is reduced to three parameters $(N, \kappa, \Gamma)$, as in the case of a macroscopic one-component plasma (OCP).
We focus on the Coulomb case with $\kappa=0$ in the following. The finite width of the peaks of the heat capacities results from the finite particle number.

The general result from Eq.~(\ref{eq:k_part_density_generalized_04}) is now applied to the Hamiltonian~(\ref{eq:h}) for the center-two-particle correlation function (C2P, $k=2$) and the triple correlation function (TCF, $k=3$) as follows:

\begin{itemize}
 \item For the C2P, we introduce spherical polar coordinates with the $z$-axis along $\vec{r}_1$, so the two 
       position vectors are ${\bf r}_1 \to \lbrace \rone, \vartheta_1, \varphi_1 \rbrace$ and 
       ${\bf r}_2 \to \lbrace \rtwo, \gangle, \varphi_2 \rbrace$, and we have $\Qr=\lbrace \rone, \rtwo, \gangle \rbrace$ 
 and $\Qi=\lbrace \vartheta_1, \varphi_1, \varphi_2 \rbrace$.
%
 In these coordinates, the angle $\gangle$ describes the angular pair distance of the two particles with respect to the trap center, cf. Fig.~\ref{fig:3d-sketch}.
 The Jacobian is $\left|\mathcal{J}\Phi\OF{\Qr,\Qi}\right| = \rone^2\sin\vartheta_1 \rtwo^2\sin\gangle$, whereas 
 the geometrical factor in the center-two-particle distribution becomes
 \begin{align}
  \vt{2}\of{\rone,\rtwo,\gangle} &= 
  \int_{0}^{\pi}{\text{d}\vartheta_1} \int_{0}^{2\pi}{\text{d}\varphi_1} \int_{0}^{2\pi}{\text{d}\varphi_2}
  \nonumber\\
  &\hspace{4em} \rone^2\sin\vartheta_1 \rtwo^2\sin\gangle
  \label{eq:c2p_vtilde_00}\\
  &= 8\pi^2 \rone^2 \rtwo^2 \sin\gangle
  \label{eq:c2p_vtilde_01} \text{ .}
 \end{align}
 Since the two-particle density must not depend on the coordinates $\Qi$, due to the rotational invariance of the Hamiltonian, we can integrate $\rho_2$ over $\Qi$ in order to obtain $\rt{2}$ in the new coordinates $\Qr$.
 The result is the product of the geometrical factor $\vt{2}\of{\rone,\rtwo,\gangle}$, Eq.~(\ref{eq:c2p_vtilde_01}), and the quantity of interest---the {\em center-two-particle correlation function}, C2P,
\begin{align}
  {\tilde \rho}_2\of{\rone,\rtwo,\gangle} = \text{C2P}\of{\rone,\rtwo,\gangle} \cdot \vt{2}\of{\rone,\rtwo,\gangle}.
\label{eq:c2p}
\end{align}

For the plots in the paper, a slightly modified C2P was chosen.
Instead of relating $\rt{2}$ to the geometrical factor $\vt{2}$, it was related to the $\rt{2}^\text{id}\of{\rone,\rtwo,\gangle} = 8\pi^{2}\rone^{2}\rtwo^{2}\rho\of{\rone}\rho\of{\rtwo}$ where $\rho\of{r}$ is the radial density. The ``ideal'' pair density $\rt{2}^\text{id}$ is the pair density that one would find in a system with the same radial density as the cluster but without any structure within the shells.
The C2P is set to unity between the shells where both $\rho_2$ and $\rho_2^\text{id}$ are zero.
\item For the TCF within a spherical shell, we introduce spherical coordinates with the $z$-axis along $\vec{r}_1$ and $y$-axis along $\vec{r}_1 \times \vec{r}_2$.
 The three position vectors are ${\bf r}_1 \to \lbrace \rone, \vartheta_0, \varphi_0 \rbrace$, ${\bf r}_2 \to \lbrace \rtwo, \gangleone, \varphi_1 \rbrace$ 
 and ${\mathbf r}_3 \to \lbrace \rthree, \gangletwo, \gangleS \rbrace$, and we have  
  $\Qr=\lbrace \rone, \rtwo, \rthree,\allowbreak \gangleone, \gangletwo, \gangleS \rbrace$ and 
  $\Qi=\lbrace \vartheta_0, \varphi_0, \varphi_1 \rbrace$. 
  To analyze intrashell order, we select three particles from the same shell and drop the radial coordinates $\rone$, $\rtwo$ and $\rthree$.   $\gangleone$ ($\gangletwo$) is the angular pair distance between the first and the second (third) particle, and $\gangleS$ is the ``bond'' angle, cf. Fig.~\ref{fig:3d-sketch}.
  The Jacobian is $\left|\mathcal{J}\Phi\OF{\Qr,\Qi}\right| = \rone^2\sin\vartheta_0 \rtwo^2\sin\gangleone \rthree^2 \sin\gangletwo$, and the geometrical factor in the triple-correlation-function on the sphere becomes
 \begin{align}
  \vt{3}\of{\gangleone,\gangletwo,\gangleS} &= 
  2 \cdot \int_{0}^{\pi}{\text{d}\vartheta_0} \int_{0}^{2\pi}{\text{d}\varphi_0} \int_{0}^{ 2\pi}{\text{d}\varphi_1}
  \nonumber\\
  &\hspace{4em} \rone^2\sin\vartheta_0 \rtwo^2\sin\gangleone \rthree^2 \sin\gangletwo
 \label{eq:tcf_vtilde_00}\\
   &= 16\pi^2 \rone^2 \rtwo^2 \rthree^2 \sin\gangleone \sin\gangletwo
  \label{eq:tcf_vtilde_01} \text{ ,}
 \end{align}
where the factor $2$ in front the integral has to be added again, since we do not resolve the orientation of the bond angles.
Since all three particles were selected from one shell with radius $\Rshell$, the volume element is $\vt{3} = 16\pi^2 \Rshell^{6} \sin\gangleone \sin\gangletwo$.
We can again split the three-particle density in the chosen coordinates into the geometrical factor and the \textit{triple correlation function}, TCF,  
\begin{align}
 {\tilde \rho}_3\of{\gangleone,\gangletwo,\phi} = {\rm TCF}\of{\gangleone,\gangletwo,\phi} \cdot 
\vt{3}\of{\gangleone,\gangletwo,\phi}.
\label{eq:tcf}
\end{align}
First applications of the TCF to spherical dust clusters were presented in Refs. \cite{schella11, thomsen_jpd14}
and confirmed that the TCF is sensitive to the gradual loss of order upon temperature increase. 
\end{itemize}

Taking advantage of the symmetries of the Hamiltonian, the reduced entropies which are associated with the two- and three-particle densities $\rho_2\of{\vec{r}_{1},\vec{r}_{2}}$ and $\rho_3\of{\vec{r}_{1},\vec{r}_{2},\vec{r}_{2}}$ are transformed into
\begin{align}
 S^{\rm (2)} \equiv - \langle \ln{\rm C2P} \rangle, \qquad 
 S^{\rm (3)} \equiv - \langle \ln{\rm TCF} \rangle \text{ ,}
\label{eq:entropies}
\end{align}
where the derivation is analogous as for the flat 2D system above, cf. Sec.~\ref{ss:2d}. 

Since the choice of $\rdelta$ has no qualitative influence on the further results, we chose $\rdelta=r_{0}$ equal to the unit of length for the C2P.
When the density $\rho_{2}$ is expressed in units of $r_{0}$, the $\rdelta$-factor is unity and it is hence omitted in the following.
For the TCF on the spherical shell, we chose $\rdelta=\Rshell$ which equals the average radius of the analyzed shell, in order to cancel the factor $\Rshell^{6}$ in the denominator that results from the volume element, cf. Eq.~(\ref{eq:three_part_entropy_generalized_02}) in Sec.~\ref{ss:2d}.

\section{Numerical simulations}\label{s:results}

\subsection{Simulation method}
We use the Metropolis Monte Carlo (MC) method where a Markov chain of configurations is generated by randomly displacing single particles. 
Each displacement is accepted or rejected with the Metropolis acceptance probability which depends on the energy change in units of $\kB T$.
One MC step describes a sequence of these displacements, so that the positions of all particles are possibly changed between two MC steps.
Since it requires a very large number of MC steps to overcome an energy barrier between two metastable configurations (``isomers'') at low temperature by chance,
the canonical ensemble cannot be sampled effectively with single particle moves only.

In order to sample all relevant isomers in reasonable simulation time, we use the parallel tempering (or exchange) Monte Carlo method~\cite{Hukushima_PaTe}. 
For this purpose, a set of 81 replica configurations is simulated in parallel for every cluster shown in this paper.
Each of these replicas has a different temperature, and the temperatures are geometrically distributed over a range that covers four orders of magnitude.
After every $M_\text{swap}=123$ MC steps a sequence of swapping moves is performed. 
For a swapping move, two adjacent replicas are chosen and their potential energies are compared.
If the hotter of the two clusters has a lower energy, then the configurations are swapped. 
If not, the swap is accepted by chance with a probability that decreases exponentially with the product of the energy difference and the difference of the inverse temperatures.
Again, accepting some of the moves which swap a configuration with a higher energy to the colder system is essential to comply with the detailed balance condition.
Having 20 replicas of the system per order of magnitude ensures the acceptance of a significant fraction of the proposed configuration swaps between adjacent temperatures.
This algorithm allows for an ergodic sampling of the accessible phase space which is verified by checking that rare isomers are equally distributed over the entire sequence of configurations even at low temperatures.

From the particle coordinates, the spatial pair and three-particle distributions functions in the generalized coordinates are sampled every $M_\text{sample}=1000$ MC steps as three-dimensional histograms.
The large number of MC steps between the samples ensures their statistical independence. Moreover $M_\text{swap}$ and $M_\text{sample}$ were chosen as prime to each other in order to exclude aliasing effects.

The reduced entropies $S^{(k)}$ are calculated after the simulation individually for each temperature point and the associated heat capacities $c^{(k)}$ are then obtained from cubic spline interpolations of $S^{(k)}$ versus $\ln\Gamma$.

\subsection{Numerical results for 3D Coulomb clusters}

\subsubsection{Typical spherical cluster $N=80$}
\begin{figure}[t]
 \includegraphics{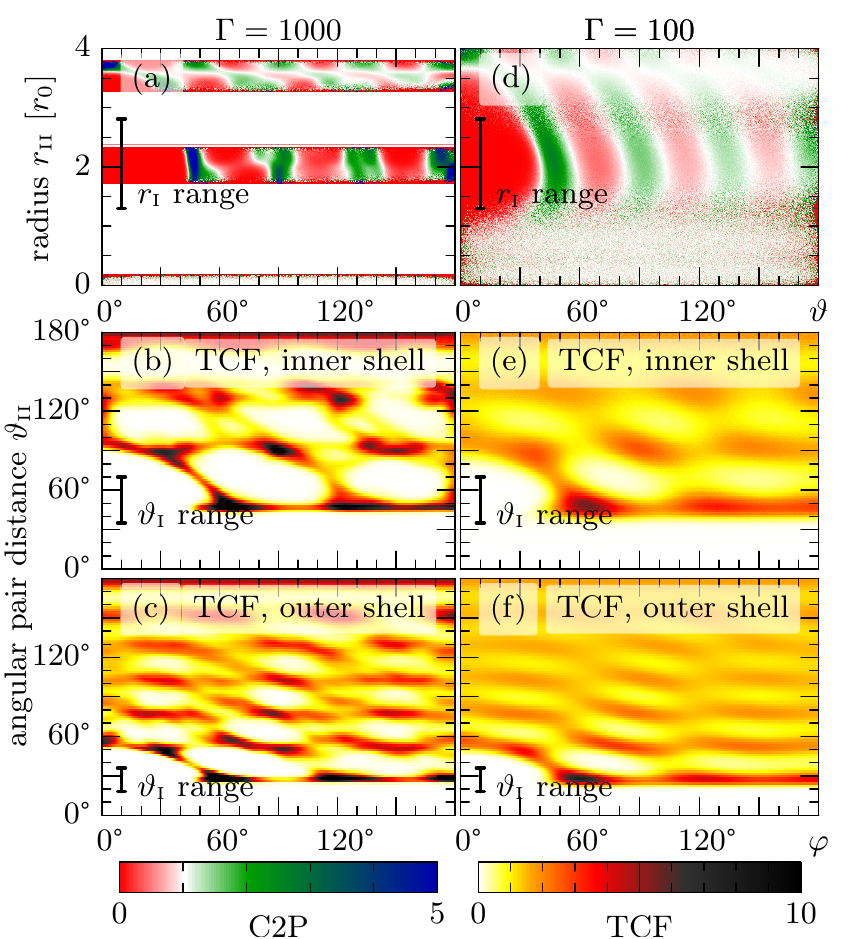}
 \caption{(Color online) Spherical 3D Coulomb cluster  with $N=80$ particles on two shells. Left (right) column: $\Gamma=1000$ ($\Gamma=100$). \textbf{(a,d):} C2P for a reference particle from the inner shell [selected by the $\rone$ integration (bar)]. \textbf{(b,e) [(c,f)]}: TCF on the inner [outer] shell. The first particle pair is selected as nearest neighbors by the $\gangleone$ integration range (bar). The length scale is $r_0$.}
 \label{fig:3d-shells_cold}
\end{figure}

We now consider, as an example, a Coulomb cluster with $N=80$ particles (Fig.~\ref{fig:3d-sketch}) which has, in the ground state, two shells with 60 and 19 particles and one particle in the trap center \cite{ludwig05}. We performed first-principle MC simulations for a $\Gamma$-range spanning four orders of magnitude.
Figure~\ref{fig:3d-shells_cold}(a) shows the C2P for $\Gamma=1000$. The first radial coordinate $\rone$ is integrated over the inner shell range (cf. the bar in the plot) to obtain $\rt{2}\of{\rtwo,\gangle}$ and $\rt{2}^\text{id}\of{\rtwo,\gangle}$, meaning that one particle is always selected from the inner shell.
Going from $\gangle={0}{\degree}$ to ${90}{\degree}$, at $\rtwo\approx 2$, a sequence of extrema is visible that reflect intrashell pair correlations whereas peaks at the radius of the outer shell, $\rtwo \approx 3.5$, 
correspond to inter-shell angular correlations.
\begin{figure}[t]
 \includegraphics{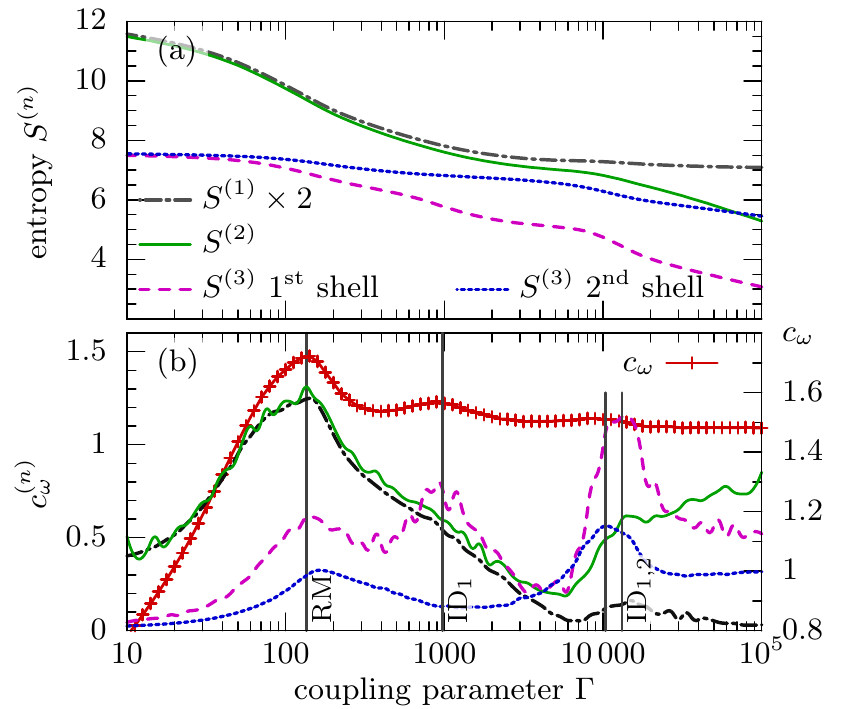}
 \caption{(Color online) \textbf{(a)} Reduced entropies for the cluster of Fig.~\ref{fig:3d-shells_cold} vs. $\Gamma$ computed for the C2P (solid line), the radial density (dash-dot) and for the TCF on the inner (dashed) and outer shell (dotted).
 \textbf{(b)} Reduced specific heat (\ref{eq:cvi_def}) for the entropies from (a) compared to the specific heat. 
RM---radial melting; ID1 (ID2)---intrashell disordering on inner (outer) shell. 
}
\label{fig:entropie_n80}
\end{figure}
The right column of Fig.~\ref{fig:3d-shells_cold} shows the same quantities for a ten times higher temperature ($\Gamma=100$)                
where all structures are completely washed out, indicating a fluid state. 

The vertical axis in the four lower parts of Fig.~\ref{fig:3d-shells_cold} shows the angular pair distances $\gangleone$ of particle one and three of the triple, whereas the horizontal axis shows the bond angle $\gangleS$, cf. Fig.~\ref{fig:3d-sketch}.
For the plots (c,f), the TCF of the outer shell was integrated over a range $\gangleone \in [{18}{\degree},{36}{\degree}]$,
which means selecting those particle triples where particle one and two are nearest neighbors.
Along the horizontal line at $\gangletwo \approx {27}{\degree}$, we find those triples where both particles two and three are nearest neighbors to the first particle.
The clear peaks around bond angles of $\gangleS = {60}{\degree}$, ${120}{\degree}$ and ${180}{\degree}$ result from a pseudo-hexagonal order within the shell, which can also be seen in Voronoi pattern in Fig.~\ref{fig:3d-sketch}.
Since the hexagons are deformed and a finite number of pentagonal Voronoi cells, i.e. particles with five nearest neighbors, exists due to the curvature of the sphere, these peaks are smeared out in $\gangleS$-direction.
Another indication for a pseudo-hexagonal intrashell structure is the $\gangleS={30}{\degree}$ peak for second neighbors.
As a ``long''-range feature of this structure at strong coupling $\Gamma=1000$ (c), we also see distinct peaks for distant particles. 
E.g., a peak at fourth neighbors ($\gangletwo\approx {102}{\degree}$) and a bond angle of $\gangleS={95}{\degree}$ reports the existence of a bond-angular order beyond nearest neighbors. When compared to the TCF of a flat 2D system, Fig.~2 (a,b), the similarities between the patterns are striking for the solid as well as for the liquid system.
Since the phase transition is much sharper in the extended system, the two selected $\Gamma$ values for solid and liquid 2D crystal only differ by a factor of two while the $\Gamma$ values for the spherical cluster differ by a factor of ten.

\begin{figure}
 \includegraphics{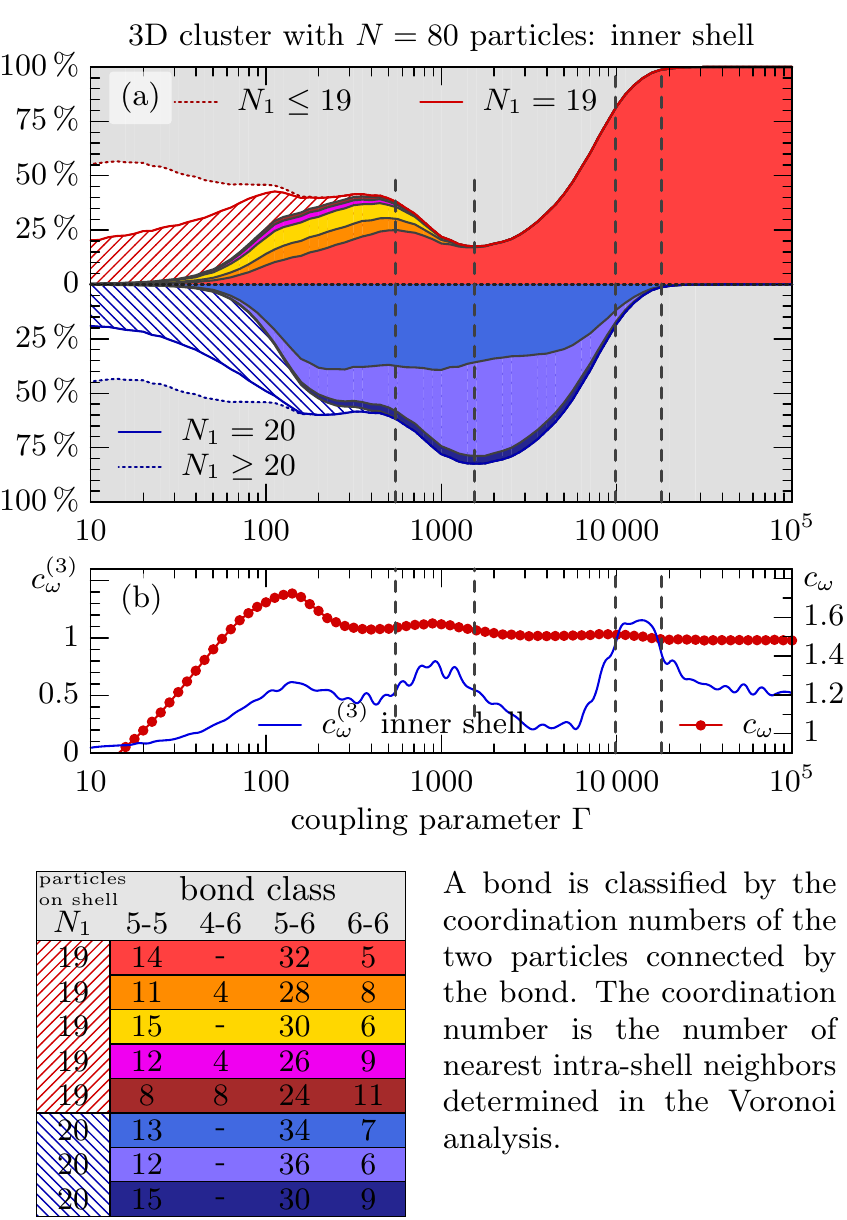}
 \caption{(Color online) \textbf{(a)} Occurrence of different states of the Coulomb cluster with $N=80$ particles. 
 The states are, first of all, divided into configurations with $N_1=19$ (solid red line, above the horizontal 0-line) particles or less (dashed) on the inner shell and those with $N=20$ (solid blue line, below the 0-line) or more (dashed). The fraction of the former (latter) configurations is plotted above (below) the zero line. 
 The intrashell order is analyzed by the means of the Voronoi construction within the shell which allows to determine intrashell nearest neighbors.
 An intrashell configuration is classified by the number of bonds of each class and the eight most important configurations plotted as filled curves. 
 The fraction of additional less important configurations with $N_1=19$ or $20$ is represented by the shaded area.\newline
 \textbf{(b)} Specific heat capacity and ``capacity'' $c^{(3)}_\omega$ from the TCF on the inner shell. The two intrashell disordering $\Gamma$-regions are indicated by vertical dashed lines.
 \label{fig:N80R1_Voro}}
\end{figure}

Due to the small number of particles on the inner shell of 19 to 20, cf. Fig.~\ref{fig:N80R1_Voro}, most particles have five instead of six nearest neighbors within the shell, here.
Hence, the TCF patterns, Fig.~\ref{fig:3d-shells_cold}(b,e), are significantly different from those of the outer shell or the flat system discussed above. E.g. the peak at a bond angle $\gangleS={120}{\degree}$ for second neighbors ($\gangleone\approx {90}{\degree}$) is missing.

To understand and quantify the details of the structural transition(s) we compute the entropies $S^{(1)}$ [from the radial density $\rho(r)$], $S^{(2)}$ (from the C2P) and two expressions for $S^{(3)}$ (from the TCF of either the inner or the outer shell). The results are shown in Fig.~\ref{fig:entropie_n80}(a) and exhibit a monotonic decrease with $\Gamma$ with several steeper drops. These again show up as distinct peaks in the derivatives (\ref{eq:cvi_def}) which we compare to the exact specific heat  $c_\omega$ in Fig.~\ref{fig:entropie_n80}(b). 
First, we notice a common peak of $c_\omega$, $c_\omega^{(1)}$ and $c_\omega^{(2)}$ around $\Gamma=136$ clearly attributed to radial melting (RM).
The nearby peaks of $c_\omega^{(3)}$ for both shells indicate loss of intrashell order at slightly higher $\Gamma$. The complete loss of order within each shell is consistent with melting observed in macroscopic 2D layers around $\Gamma=137$, e.g.~\cite{hartmann_2d,ott_pop}. Another peak in $c^{(3)}_\omega$ [pink dashed line (blue dotted line)] is seen at $\Gamma\approx 13\,156$ ($\Gamma\approx 11\,000$) indicating the onset of intrashell disordering within the inner (outer) shell, due to transitions between different configurations of Voronoi pentagons and hexagons (``intrashell isomers'') \cite{ludwig05}.
This transition is accompanied by the transition between two radial isomers. While at high coupling, the outer shell is always occupied by $N_{2}=60$ particles, we observe a significant probability of occurence of the configuration with $N_{2}=59$ particles on this shell, for lower $\Gamma$.

Finally, there is another peak in $c^{(3)}_\omega$ of the inner shell (and in $c_\omega$) around $\Gamma=977$  which is associated with enhanced transitions of particles between shells \cite{intershell_trans}. 
By means of the intrashell Voronoi analysis, we find that, for $N_{1}=19$, the shell configuration splits up into several intrashell isomers around this value of $\Gamma$. 
While for $N_{1}=20$ particles on the inner shell different intrashell isomers are found also at higher coupling, we find only one intrashell configuration at $\Gamma \gtrsim {1000}$ when this shell is occupied by $N_{1}=19$ particles.
%
%

\subsubsection{Magic spherical cluster $N=38$}
\begin{figure}
 \includegraphics{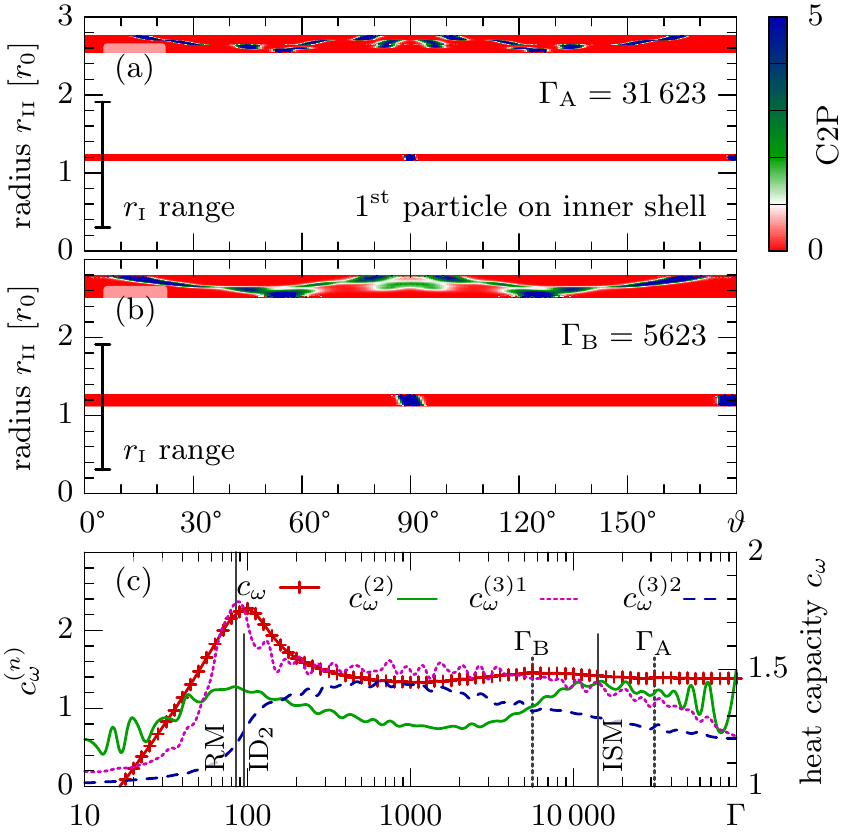}
 \caption{(Color online) Inter-shell angular melting (ISM) in the spherical 3D Coulomb cluster with $N=38$ particles. \textbf{(a),(b)} C2P   [the first particle radius is averaged over the inner shell] below (above) the melting temperature. \textbf{(c):} 
 specific heat and reduced specific heat vs. $\Gamma$. While radial melting  (RM) is seen in $c_\omega$ and $c_\omega^{(2)}$, ISM is clearly visible only in $c^{(2)}_\omega$, based on the C2P.}
\label{fig:c2p_38}
\end{figure}
The results shown above are typical for finite spherical Coulomb and Yukawa clusters except for ``magic'' clusters which exhibit particular stability against certain excitations. In particular, we inquire whether an \textit{inter-shell angular melting} (ISM) transition, known from 2D clusters \cite{bedanov,filinov-etal.01prl}, exists here as well.
As an example we consider the case $N=38$ with 6 (32) particles on the inner (outer) shell and study the C2P for particle pairs from two different shells.
The associated inter-shell angular correlations are visible in Fig.~\ref{fig:c2p_38}(a) as localized spots in the horizontal strip around $\rtwo=2.7$. At a five times higher temperature, cf. Fig.~\ref{fig:c2p_38}(b), these spots overlap, indicating ISM. This is confirmed by the entropy $S^{(2)}$ computed from the C2P and the associated specific heat $c_\omega^{(2)}$ [Fig.~\ref{fig:c2p_38}(c)] which exhibits a clear peak around $\Gamma=14\,175$. 
Our interpretation is confirmed by the missing of this peak in $c_\omega^{(3)}$ [computed from the TCF for particles on the same shell]. Interestingly, also the full specific heat $c_\omega$ misses this peak \cite{more_peaks_38} which indicates that the reduced quantities $c_\omega^{(k)}$---when computed for adequately selected particle pairs or triples---may be even more sensitive to structural transitions than the full heat capacity.

For the ``magic'' particle number $N=38$, the intrashell order freezes together with the radial structure.
Due to the high symmetry of this order, a high temperature is required to find different intrashell isomers. 
The reduced heat capacity $\cheat^{(3)1}$ of the TCF on the inner shell ($N_{1}=6$) agrees with the radial melting peak in the heat capacity.
Interestingly, for the outer shell $\cheat^{(3)2}$ shows a step-like rise at this point rather than a peak.
During their numerical study, Calvo and Yurtsever found two different isomers with the same (6, 32) radial composition for $\Gamma\lesssim 20\,000$ with very close energy ($\Delta E=3\cdot 10^{-4}$), whereas only one isomer was found, at higher coupling~\cite{calvo_melting_2007}.
Our interpretation is that these two isomers differ with respect to the relative orientation of inner and outer shell and the occurrence of more than one isomer indicates the inter-shell angular disordering process which we observe by means of the C2P at approximately the same coupling strength.
In Sec.~\ref{sse:ADF}, we compare our findings from $\cheat^{(k)}$ with the fluctuations of the angular distance between adjacent particles as a Lindemann-like melting parameter for this cluster~\cite{apolinario_2007}. 

\subsubsection{Larger spherical cluster $N=120$}
For the two investigated Coulomb clusters with $N=80$ and $N=38$ particles, we found intrashell disordering processes which were accompanied by a transition between two radial isomers or by radial melting processes, respectively.
This raises the question whether intrashell disordering can take place without a change of the particle number in the shell, for certain 3D clusters.

\begin{figure}
 \includegraphics{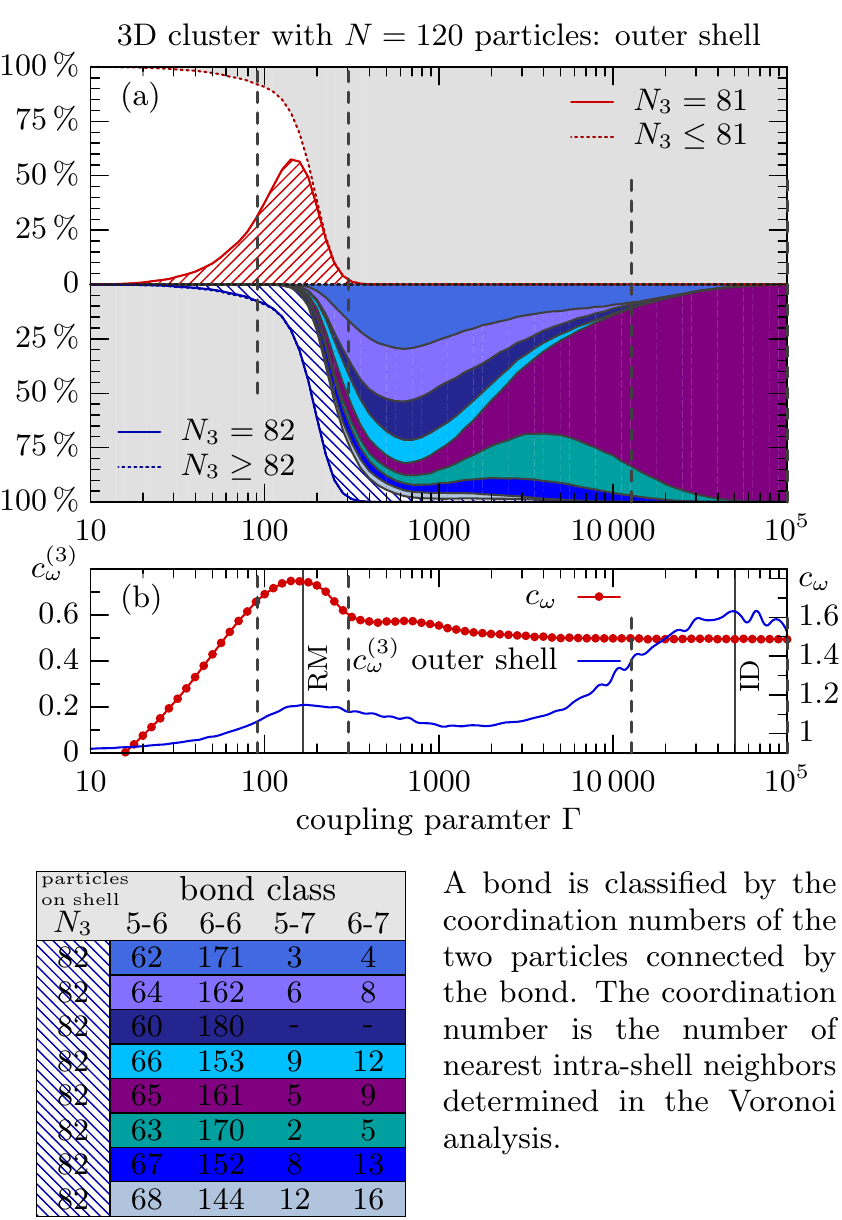}
 \caption{(Color online) Intrashell disordering transition (ID) in the outer shell of a spherical Coulomb cluster with $N=120$ particles around $\Gamma=50\,000$.
 \textbf{(a)} The particle number on the outer shell is constant for $\Gamma\gtrsim 400$ while transitions between different intrashell isomers take place at higher $\Gamma$.
 The intrashell Voronoi analysis allows for a classification of these isomers.
 \textbf{(b)} While the thermodynamic heat capacity $\cheat$ captures only the radial melting (RM) at $\Gamma\approx165$, the reduced heat capacity $\cheat^{(3)}$ computed from the entropy of the TCF captures also the intrashell disordering.
 \label{fig:N120R3_Voro}}
\end{figure}

Such a transition is, indeed, found in a larger Coulomb cluster with $N=120$ particles. 
The ground state configuration is (82, 32, 6), counting the particle numbers from outer to inner shell. 
Up to $\Gamma\approx 400$, no isomers with different shell occupation numbers are found in the simulations although the heat capacity $\cheat^{(3)}$ shows a peak already at $\Gamma\approx50\,000$, see Fig~\ref{fig:N120R3_Voro}(b). 
This capacity was calculated from the TCF on the outer shell and it is hence sensitive to transitions within that shell.
We performed an intrashell Voronoi analysis in order to elucidate this low temperature transition.
As shown in Fig.~\ref{fig:N120R3_Voro}, different intrashell isomers make up a significant fraction of the ensemble, for lower coupling strength, while virtually only one isomer is found for higher coupling strength. Interestingly this isomer has two particles with seven nearest neighbors in the Voronoi cell.
The analysis of the outer shell's structure by means of the Voronoi diagram confirmed the assumptions that the peak in $\cheat^{(3)}$ (outer shell) is connected to an intrashell disordering transition without any effect on the radial structure.

\subsection{Comparison with angular distance fluctuations\label{sse:ADF}}
In this section, we compare our novel melting parameters with the mean angular distance fluctuations (ADF) for the magic number clusters with $N=38$ particles. 
The ADF are a Lindemann-like parameter that has been frequently use to analyzed the inter-shell fluctuations in spherical 2D clusters, e.g.~\cite{bedanov, filinov-etal.01prl}. 
More recently the ADF have been applied also to the intrashell and intershell fluctuations of 3D Coulomb clusters by Apolinario \textit{et al.}~\cite{apolinario_2007}.
The ADF provide a qualitatively different approach to detecting transition processes in finite systems. 
The ADF captures the property of the solid phase that the neighbors of any particle remain the same over a very long time.
In contrast, in the fluid phase, the neighbors change frequently. This means, in the solid phase, the angular distance of neighboring particles exhibits small oscillations whereas the amplitude becomes large as particles leave their neighborhood in the fluid phase.
The ADF are defined by~\cite{apolinario_2007}
\begin{align}
 \Delta \alpha_{\gamma\beta} &= \frac{1}{N_{\gamma}} \sum_{i=1}^{N_{\gamma}} \langle \alpha^{2}_{ij} \rangle - \langle \alpha_{ij} \rangle^{2}
 \label{eq:ADF_def} \text{ ,}
\end{align}
and measure the angular displacement for a pair of particles within a given shell ($\gamma=\beta$) or from different shells ($\gamma \neq \beta$).
The particle number on shell $\gamma$ is denoted by $N_{\gamma}$ and $\alpha_{ij}$ is the angular pair distance with respect to the trap center between particle $i$ on shell $\gamma$ and its nearest neighbor $j$ on shell $\beta$~\cite{apolinario_2007}. 

There are several methods to evaluate Eq.~(\ref{eq:ADF_def}) which differ in the number of neighbors included in the sum and in the simulation scheme. Here we use Langevin molecular dynamics (LMD) simulations and two versions. 
In the first variant denoted by $\Delta\alpha_{\gamma\beta}^\text{closest}$, the nearest neighbor $j$ of a given particle $i$ is determined once, at the beginning of the simulation.
In a second variant denoted $\Delta\alpha_{\gamma\beta}^\text{all}$, all pairs of particles $ij$ from shells $\gamma$ and $\beta$ are sampled, disregarding their proximity.
Both variants are calculated in LMD simulations using a time step $\Delta t=0.005$ and a damping coefficient $\gamma=0.5$, in units of $t_{0}$ and $t_{0}^{-1}$, respectively. 
The LMD simulations were initialized by a configuration from the MC simulation at high coupling strength $\Gamma$, and $10^{6}$ time steps were performed for data production after the equilibration.

\begin{figure}
 \includegraphics{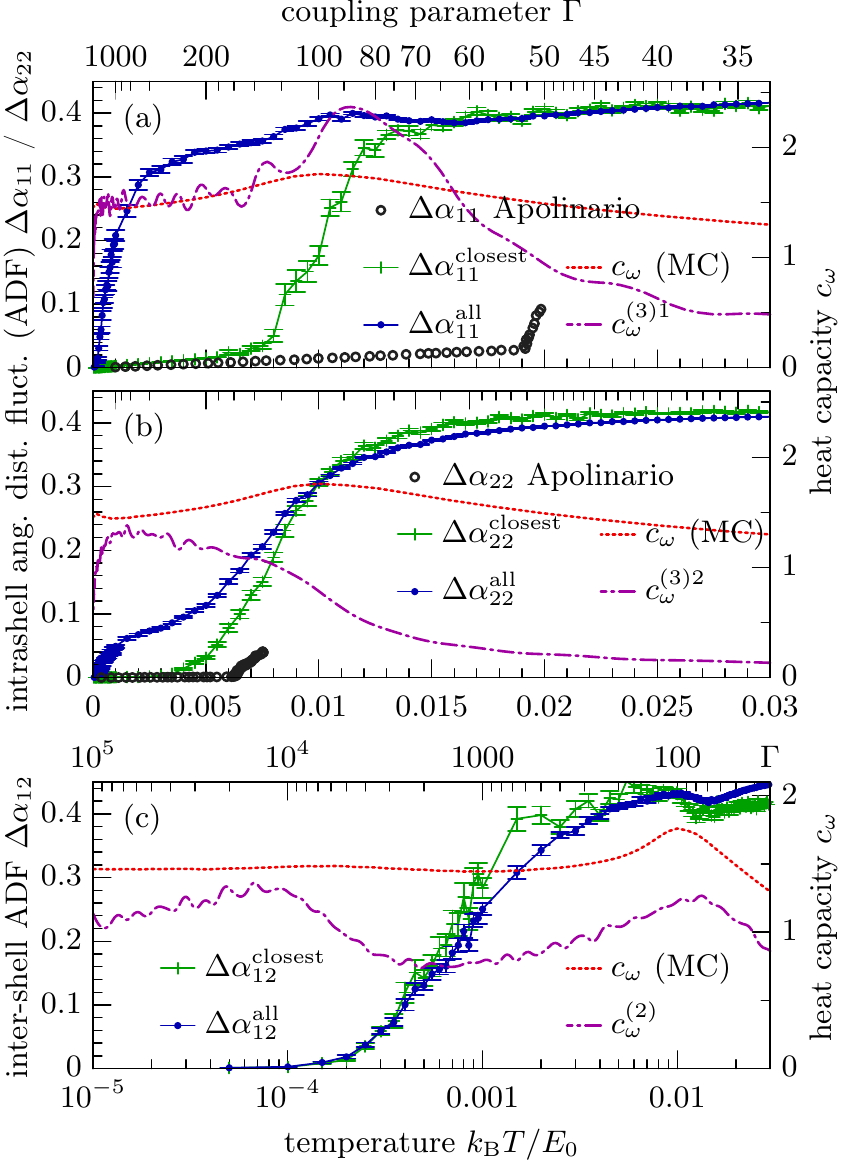}
 \caption{(Color online) Comparison of reduced heat capacities, $\cheat^{(k)}$, and angular distance fluctuations (ADF), $\Delta\alpha$, Eq.~(\ref{eq:ADF_def}), for the magic number Coulomb cluster with $N=38$ particles.
  \textbf{(a)} Intrashell disordering on the inner shell, measured by $\cheat^{(3)1}$ from the TCF and by $\Delta\alpha_{11}$  for pairs of two particles within the inner shell.  \textbf{(b)} The same for the outer shell.
  \textbf{(c)} Inter-shell angular disordering, measured by $\cheat^{(2)}$ from the C2P and by  $\Delta\alpha_{12}$, for pairs of particles belonging to different shells.
  While the (reduced) heat capacity $\cheat$ ($\cheat^{(k)}$) was obtained in an MC simulation with parallel tempering, the fluctuations $\Delta\alpha$ were obtained as averages over 40 Langevin molecular dynamics (LMD) simulation runs for each temperature. 
In Figs.~(a,b) we also compare to data by Apolinario \textit{et al.} contained in Fig.~5 of Ref.~\cite{apolinario_2007}. (The temperature was scaled by $\sqrt[3]{0.5}$ because of the different energy unit used in that reference.) Note the different temperature scale in Fig.~(c).
 \label{fig:3DN038_lindemann}}
\end{figure}

Figure~\ref{fig:3DN038_lindemann}(a) shows the ADF for the six particles on the inner shell of the cluster.
The sharp increase of $\Delta\alpha^\text{closest}_{11}$ is in good agreement with the peak in the reduced heat capacity $\cheat^{(3)1}$ from the TCF on that shell.
Part (b) of the figure refers to the outer shell. The increase of $\Delta\alpha^\text{closest}_{22}$ takes place around $\Gamma=100$. Around this point, the reduced heat capacity has a broad peak. The increase of $\Delta\alpha^\text{closest}_{22}$ 
falls on the decreasing slope of this peak
which is seen more clearly in Fig.~\ref{fig:c2p_38}(c). [Note that in Fig.~\ref{fig:c2p_38}(c) the $\Gamma$-axis is inverted.]

Finally, Fig.~\ref{fig:3DN038_lindemann}(c) shows the inter-shell ADF, in comparison with $\cheat^{(2)}$ obtained from the C2P function.
In view of the low critical temperature for the loss of the inter-shell angular order observed in Fig.~\ref{fig:c2p_38} above, we chose a logarithmic temperature scale for this figure part.
While the peak of $\cheat^{(2)}$ indicates that the inter-shell angular disordering transition occurs around $\Gamma_\text{ISM}={14\,175}$, the increase of the ADF $\Delta\alpha_{12}$ is found at significantly lower coupling strength (higher temperature). This discrepancy is interesting but not fully understood yet.
%
While the C2P function depends on the accessible phase space volume, the ADF are sensitive to the evolution of the cluster.
During this evolution, the transition between two isomers which differ with respect to the relative angle of the two shells constitutes a rare event, at low temperatures.
Hence the LMD simulation time can become insufficient to obtain good statistics for these events, in the very high coupling regime.

As the C2P function resolves also the radial structure of the dust cluster, $\cheat^{(2)}$ exhibits a second peak at the radial melting point, around $\Gamma_\text{RM}={85}$, cf. Fig.~\ref{fig:c2p_38}(c), which has no equivalence in the angular distance fluctuations.

In Figs.~\ref{fig:3DN038_lindemann}(a,b) we also compare with data by Apolinario and Peeters (open black dots), for the same cluster~\cite{apolinario_2007}, cf. Fig. 5 of that reference. In all our simulations the increase of $\Delta\alpha$ is found to occur at significantly lower temperatures. This may result from a different treatment of neighbors during the calculations of the ADF or from differences in the molecular dynamics simulations.
\newline

\section{Conclusion}\label{s:conclusion}
In summary, we have proposed novel quantities for the analysis of structural transitions in inhomogeneous finite systems---the reduced entropies $S^{(k)}$ and the associated reduced heat capacities $\cheat^{(k)}=T\,\pd{S^{(k)}}{T}$ computed from the $k$-particle distribution functions. Our results indicate that spherical 3D clusters with long range interaction \cite{yukawa_results} exhibit (at least) three different structural transitions: i) inter-shell angular melting (at $\Gamma \gtrsim 10^4$), ii) radial melting, at $\Gamma \sim140$ and iii) intrashell disordering which starts, in different shells at different $\Gamma$ around $10,000$ and typically extends up to the RM transition.

We note that different melting processes in 3D Coulomb clusters were investigated before, for a detailed study, see Ref.~\cite{calvo_melting_2007}.
These authors performed both parallel tempering MC simulations to access the specific heat and molecular dynamics (MD) simulations to obtain dynamic properties, such as the Lindemann parameter.  
The analysis of the spatial pair- and three-particle correlation functions performed in the current work allowed us to 
obtain complementary information. In particular, we are now able to distinguish between radial melting, inter-shell angular melting and intrashell transitions.
Since these functions are thermodynamic properties of the system, they can be obtained in the same MC simulation as the thermodynamic heat capacity and require no knowledge about the dynamic properties. Moreover, these quantities are very well compatible with enhanced Monte Carlo techniques such as parallel tempering. 

The reduced entropies $S^{(k)}$ and the reduced heat capacities $c^{(k)}$ are easily computed and directly measurable in experiments with colloids or dusty plasmas where the particle positions are detected, and it will be interesting to verify the reported structural transitions.

In the present work, we were able to assign the additional peaks in $\cheat^{(k)}$ to certain disordering processes. 
An open question is the connection between these and phase transitions of the cluster and whether one can determine the order of these transitions, in analogy to macroscopic systems.

Finally, the proposed method of reduced entropies and heat capacities is not restricted to spherical traps. It should be equally applicable to other finite systems, including interfaces, as well as to quantum systems.

This work is supported by the Deutsche Forschungsgemeinschaft via SFB-TR 24 (project A9) and by CPU time at the HLRN (grant SHP006).


\begin{thebibliography}{26}
\bibitem{intershell_trans} 
 For $\Gamma \lesssim 6000$ the shell configuration (59,20,1) dominates, but around $\Gamma=1000$ there is a strong reentrance of the configuration (60,19,1), see Fig.~\ref{fig:N80R1_Voro}.
\bibitem{yukawa_results}
 While the present results have been obtained for Coulomb clusters the behavior of Yukawa clusters is similar. An extensive study of the screening dependence of the structural transitions is beyond the range of this paper and will be given elsewhere.
\end{thebibliography}
\end{document}